\documentstyle[11pt,graphicx]{article}
\title{An Analytic Approach to the CMB Polarization
Generated by Relic Gravitational Waves   }
\author{\small   Wen  Zhao and   Yang  Zhang \\
        \small Astrophysics Center \\
       \small University of Science and Technology of China \\
       \small Hefei, Anhui, China }
 \date{}

\begin{document}
\maketitle
\baselineskip=19truept
\def\vek{\vec{k}}

\newcommand{\be}{\begin{equation}}
\newcommand{\ee}{\end{equation}}
\newcommand{\ba}{\begin{eqnarray}}
\newcommand{\ea}{\end{eqnarray}}

\sf
\small

\begin{center}
\Large  Abstract
\end{center}
\begin{quote}
 {
We develop Polnarev's analytic method to calculate of the
polarization power spectrum of the cosmic microwave background
radiation (CMB) generated by cosmic relic gravitational waves. In
this analytic approach the physics involved in this process is
more transparent. Consequently, the effects due to various
elements of physics can be isolated easily. In numerically
calculating the evolution of the gravitational waves during the
radiation-matter transition, the WKB approximation for the scale
factor has been taken.
To describe more precisely the decoupling
process, we have introduced an analytic expression for the
visibility function, consisting of two pieces of half-gaussian
curves.
We also include the damping effects on the power spectrum
at small scale up to the second order
of the tight coupling due to the collisions.
An analytic polarization spectra $C_l^{XX}$ has been obtained
with following several improvements over the previous results.
1. The approximate analytic  result is quite
close to the numerical one evaluated from the cmbfast code,
especially, for the first three peaks of the spectrum that are
observable. By using the analytic exact solution of relic gravitational waves
 from the sudden-change approximation, we have demonstrated the
dependence of  $C_l^{XX}$ on the dark energy and the baryon.
2. Our analytic half-gaussian approximation of the visibility function
fits analytically better than the usual Gaussian model,
and its time integration yields a parameter-dependent damping factor.
This improves the
spectrum $\Delta C_l^{XX}/ C_l^{XX} \sim 30\%  $ around the second
and third peaks.
 3. The second order of tight coupling limit
reduces the amplitude of $C_l^{XX}$ by $58\%$, comparing with the
first order.
4.  The influence of the power spectrum index of relic GW
is such that a larger value of $n_T$ produces
higher polarization spectra $C_l^{XX}$.

 }
\end{quote}



e-mail: yzh@ustc.edu.cn

\newpage
%\twocolumn
\baselineskip=21truept

\large

\begin{center}
{\em\Large 1.~~ Introduction.}
\end{center}

Studies on the anisotropy and polarization of CMB have made great
progress, yielding important information of cosmology. Recently,
Wilkinson Microwave Anisotropy Probe (WMAP) observational results
on the CMB anisotropy and polarization power spectrums
\cite{map1-over,map1-te,map1-para,map1-inflation,map3-over,map3-pola}
agree well with the prediction of inflation of a spatially flat
Universe with the nearly scale-invariant and Gaussian spectrum of
primordial adiabatic perturbations. Inflationary expansion can
generate two type of perturbations $h_{ij}$ of spacetime metric:
one is the scalar type (density) of perturbations
\cite{sasaki,tensor1, muk}, and the other is the tensorial type
(relic gravitational waves)
\cite{Rubakov:1982,Fabbri:1983,Abbott:1984,Starobinskii:1985,Polnarev:1985,sasaki,muk}.
These two kinds of perturbations will enter the Boltzmann equation
for photons and influence  CMB during the decoupling. Their impact
on the anisotropy and polarization of CMB are different,
especially their respective contributions are not completely
determined theoretically. The major contribution to CMB
anisotropies and polarization are from the density perturbations
\cite{sz,peeble}. However, the tensorial contribution is also
important, especially in long wave-length range. Moreover, the
magnetic type of polarization of CMB can only be generated by the
tensorial perturbations, and it thus provides another channel to
detect the relic gravitational waves besides the direct detection
of laser interferometer \cite{ligo,lisa, bbo,decigo}.

The power spectra of CMB polarization can be calculated by
numerical method \cite{cmbfast,camb}, which gives rather precise
predictions. But the semi-analytic method is also very helpful in
analyzing the underlying   physics and in revealing the dependence
on the cosmological parameters \cite{Hu}. A common treatment uses
the spherical harmonic functions to expand the Boltzmann equation
into a hierarchical set of equations for the multipole moments,
then solves each of them separately \cite{cmbfast,zalda, grish,
prit}. The other treatment was first suggested by Polnarev
\cite{Polnarev:1985}, using a basis of polarization vectors to
decompose the Boltzmann equation, ending up with only two
equations for the two unknown distribution functions, $\zeta$ and
$\beta$, standing for the anisotropy and the polarization,
respectively. This treatment is simpler and has been further used
\cite{harari,Ng,koso,kami,keat,cabe,zhz}. In this paper, we study
the CMB polarization caused by the relic gravitational waves in
the Polnarev framework. Our result is an analytic formula of the
polarization power spectrum, which depends explicitly on the
visibility function and on the spectrum of the relic gravitational
waves at the decoupling. In our treatment for the ionization
history through recombination we use a half-gaussian visibility
function, which is more precise than the usual gaussian function.
For the relic gravitational waves we use  a WKB approximation.
When doing integration for the Boltzmann equations, we find there
are two kind of damping mechanism on the power spectrum: one is
caused by the visibility function, which causes damping at small
scale, and another is the second order tight coupling limit when
solving the Boltzmann equations, which is only a simple factor
damping in all scale. The latter is different from the case of the
scalar perturbation sources, where the so-called "silk damping",
which only causes the damping for the power spectrum at the small
scale. We also find that the finial power spectra depend
sensitively on the state of tensor perturbations at the decoupling
time (the ratio of the positive and negative modes), which is
complex and prevents us from getting an exact formula for the
power spectrums.

The organization of this paper is as follows:  in Section 2 and 3,
we write down the evolution equations of the polarization
distribution function $\beta_l$, and the exact relations of
$C_{l}^{GG}$ and $C_{l}^{CC}$ with it. For the following
discussion, we calculate the evolution of the gravitational waves
and build the visibility function models in Section 4 and 5. In
Section 6, we have an approximately analytic calculation of the
Boltzmann equations, and give the expression of $\beta_l$, which
follows the expression of the power spectrum functions
$C_{l}^{GG}$ and $C_{l}^{CC}$. At last, we give a conclusion and
discussion in Section 7, where we main discuss the effect of
tensor-scalar ratio $r$, the baryon density $\Omega_b$ ,dark
energy density $\Omega_{\Lambda}$ and the reionization process on
the power spectrums.

~

\begin{center}
{\em\Large 2.~~ CMB Polarization.}
\end{center}

The polarized distribution function of photons is generally represented by a
column vector $f=(I_l, I_r, U, V)$, and
its components are related to the Stokes parameters:
 $I= I_l+I_r$ and $Q=I_l-I_r$.
An important property of the Stokes parameters is that,
under a rotation $\delta$ about the axis of propagation,
the total intensity $I$ and the parameter $V$ are invariant,
but $Q$ and $U$ transform as \cite{chan}
\[
\left(
 \begin{array}{c}
 Q'\\
 U'
  \end{array}
 \right)
 =
\left( \begin{array}{cc}
\cos 2\delta & \sin 2\delta \\
-\sin 2\delta & \cos 2\delta
 \end{array}
 \right)
\left(
 \begin{array}{c}
 Q\\
 U
  \end{array}
 \right).
\]
So $(Q, U)$ together form a spin-$2$ field according to the
coordinate transformation,
and can be conveniently described by a $2\times2$ polarization tensor.
For actual detections the photon come from the full sky of 2-sphere,
which in the spherical coordinates $(\theta, \phi$) has  a metric
\begin{equation}  \label{metric}
g_{ab}= \left(
\begin{array}{cc}
1 & 0\\
0 & \sin^2\theta\\
\end{array}
\right),
\end{equation}
the polarization tensor is \cite{kami}
\begin{equation} \label{tensor}
P_{ab}(\hat{n})= \frac{1}{2} \left(
\begin{array}{cc}
Q(\hat{n}) & -U(\hat{n})\sin\theta\\
-U(\hat{n})\sin\theta & Q(\hat{n})\sin^2\theta\\
\end{array}
\right),
\end{equation}
satisfying $P_{ab}=P_{ba}$, and $g^{ab}P_{ab}=0$.

During the era prior to the decoupling in the early Universe,
the Thompson scattering of anisotropic radiation by free electrons can
give rise to the linear polarization only,
and does not generate the circular polarization $V$,
so we only consider $f=(I_l, I_r, U)$.
For the trivial case of a homogeneous and isotropic unpolarized radiation,
the distribution is simply
$f=f_0(\nu)(1,1,0)$,
where $  f_0(\nu)=\frac{1}{e^{h\nu/kT}-1} $
is the usual blackbody distribution function with temperature $T$.
The combined effects of the Thompson scattering and the metric
perturbations will yield linear polarizations of photons.
The time evolution
of the photon distribution function is determined by the equation
of radiative transfer, essentially the Boltzmann
equation \cite{chan},
 \be \label{boltzeq}
 \frac{\partial f}{\partial\eta}+\hat{n}^i
\frac{\partial f}{\partial x^i}= -\frac{d
\nu}{d\eta}\frac{\partial f}{\partial \nu}
-q(f-J),\label{boltzmamm}
 \ee
where $\hat{n}^i$ is the unit vector in the direction $(\theta,
\phi)$ of photon propagation, $q=\sigma_T n_e a$ is the
differential optical depth and has the meaning of scattering rate,
$\sigma_T=6.65\times 10^{-25}cm^{-2}$ is the Thompson
cross-section, $n_e$ is the number density of the free electron,
and
\be
 J=\frac{1}{4\pi}\int_{-1}^{1}d\mu'\int_{0}^{2\pi}d\phi'
 P(\mu,\phi,\mu',\phi')f(\eta,x^i,\nu,\mu',\phi'),\label{J}
 \ee
where $\mu=\cos\theta$, $\mu'=\cos\theta'$ and
 \be
 P=\left(
\begin{array}{ccc}
     \mu^2\mu'^2\cos2(\phi'-\phi) & -\mu^2\cos2(\phi'-\phi) &
     \mu^2\mu' \sin2(\phi'-\phi)\\
     -\mu^2\cos2(\phi'-\phi)&\cos2(\phi'-\phi)&-\mu\sin2(\phi'-\phi)\\
     -2\mu\mu'^2\sin2 (\phi'-\phi) & 2\mu \sin2(\phi'-\phi) & 2
     \mu\mu'\cos2(\phi'-\phi)
\end{array}
\right)
 \ee
is the phase-matrix.
The scattering term $q(f-J)$  in Eq.(\ref{boltzeq})
describes the effect of the
Thompson scattering by free electrons, and the term
$-\frac{d\nu}{d\eta}\frac{\partial f}{\partial \nu}$
reflects the effect of
variation of frequency due to the metric perturbations
through  the Sachs-Wolfe formula \cite{sa}
\[
\frac{1}{\nu}\frac{d\nu}{d\eta}
= \frac{1}{2} \frac{\partial
  h_{ij} }{\partial \eta}\hat{n}^i \hat{n}^j.
\]
In the presence of perturbations $h_{ij}$,
either scalar  or  tensorial,
the distribution function  will be perturbed and can be
generally written as
 \be
 f(\theta,\phi)= f_0\left[
 \left(
 \begin{array}{c}
 1\\
 1\\
 0\\
 \end{array}
 \right) +f_1\right] ,
 \ee
where  $f_1$  represents the perturbed portion.

The perturbed flat Robertson-Walker (FRW) metric is
 \be
 ds^2=a^2(\eta) \left[  d\eta^2-(\delta_{ij}+h_{ij})dx^idx^j \right]. \label{metric}
 \ee
where $\eta=\int (a_0/a)  dt$ is the conformal time,
and $h_{ij}$ are the perturbations  with $|h_{ij}| \ll 1$.
In our context, we consider only the tensorial type perturbations $h_{ij}$,
representing the relic gravitational waves.
So they are symmetric $h_{ij}=h_{ji}$, traceless $h_{ii}=0$,
and transverse $h_{ij,j}=0$.
Therefore,  there are only two independent modes,
corresponding to the $+$ and $\times$ gravitational-wave polarizations.
\[
h_{ij} = h^{+}_{ij}+ h^{\times}_{ij} = h^{+} \epsilon^+_{ij}
           + h^{\times}\epsilon^{\times}_{ij}.
 \]
Taking the direction of propagation of the GW
in the direction of $\hat z$, $\hat{k}=\hat{z}$,
then the polarization tensors for the
GW satisfy
\[
\epsilon^{+}_{ij}\hat{n}_{i}\hat{n}_j  =\sin^2 \theta \cos 2\phi,
\,\,\,\,\,\,\,\,\,\,\,
\epsilon^{ \times }_{ij}\hat{n}_{i}\hat{n}_j =\sin^2 \theta \sin 2\phi.
\]
In cosmological context, it is usually assumed that the two
components $h^+$ and $h^{\times}$ have the same magnitude and are
of the same statistical properties.
To simplify  the Boltzmann equation (\ref{boltzmamm}),
for the $h_{ij} = h^{+} \epsilon^{+}_{ij}$ polarization,
one writes the perturbed
distribution function  $f_1$ in the form \cite{Polnarev:1985}
 \be \label{f1+}
 f_1= \frac{\zeta }{2} \left(1-\mu^2\right)\cos2\phi
 \left( \begin{array}{c}
1\\
1\\
0
\end{array}
\right)
 + \frac{\beta }{2} \left(
\begin{array}{c}
(1+\mu^2)\cos2\phi\\
-(1+\mu^2)\cos2\phi\\
4\mu\sin2\phi
\end{array}
\right).
 \ee
For the $h_{ij} = h^{\times} \epsilon^{\times}_{ij}$ polarization,
one writes $f_1$  in the form
  \be
 f_1= \frac{\zeta}{2} \left(1-\mu^2\right)\sin2\phi
 \left( \begin{array}{c}
1\\
1\\
0
\end{array}
\right)
 + \frac{\beta }{2} \left(
\begin{array}{c}
(1+\mu^2)\sin2\phi\\
-(1+\mu^2)\sin2\phi\\
 - 4\mu\cos2\phi
\end{array}
\right),
 \ee
where $\zeta$ represents the anisotropy of photon  distribution
since $\zeta\propto I_l +I_r = I$,  and $\beta$ represents the
polarization of photons since $\beta \propto  I_l -I_r =Q$.
Both $\zeta$ and $\beta$ are to be determined by solving the Boltzmann equation.
For the $h_{ij} = h^{+} \epsilon^{+}_{ij}$ polarization,
one substitutes $f$ into Eq.(\ref{boltzmamm}).
Upon  taking Fourier transformation,
retaining only the terms linear in the perturbation $h_{ij}$,
and performing the integration over $d\mu$,
one arrives at a set of two differential equations \cite{Polnarev:1985,cabe, zhz},
\begin{eqnarray} \label{eqn:rewritten1}
   \dot{\xi}_k+\left[ik\mu+q\right]\xi_k  =
      \frac{d \ln f_0 }{d \ln \nu_0}    \dot{h}^+_k ,
\end{eqnarray}
\begin{eqnarray}
     \dot{\beta}_k+\left[ik\mu+q\right]\beta_k  =  \frac{3}{16} q\int
     d\mu'\left[\left(1+\mu'^2\right)^2\beta_k -\frac{1}{2}\left(1-\mu'^2\right)^2\xi_k\right].
     \label{eqn:rewritten}
\end{eqnarray}
For  the $h_{ij} = h^{\times} \epsilon^{\times}_{ij}$ polarization,
the resulting equations are
the same as Eqs.(\ref{eqn:rewritten1}) and (\ref{eqn:rewritten})
with $\dot h^+$ being replaced by $\dot h^\times$.
In Eq.(\ref{eqn:rewritten1}) and (\ref{eqn:rewritten})
$\xi_k  \equiv \zeta_k +\beta_k $,
$k$ is  the wavenumber,
$\xi_k$, $\beta_k$, and $h_k^+$ are the Fourier modes of
$\xi$, $\beta$, and $h^+$, respectively,
and the over dot $``\cdot"$ denotes $d/d\eta$.
In the following, for simplicity,
we will omit the sub-index $k$ of the function $\xi_k$,
$\beta_k$ and $h_k$.
Moreover, we also drop the GW polarization notation, $+$ or $\times$,
since both $h^+$ and $h^\times$ are similar in computations.
From the structure of Eqs.(\ref{eqn:rewritten1}) and (\ref{eqn:rewritten}),
one can see
that the $ \dot{h} $ of GW in Eq.(\ref{eqn:rewritten1})
plays the role of a source for the anisotropies $\xi$,
which in turn plays the role of a source for the polarization $\beta$
in Eq.(\ref{eqn:rewritten}).
Our work  is to
find the solution of $\beta$,
then calculate the CMB polarization power spectra.

In the simple case of the long-wavelength limit with $k=0$,
this set of equations  reduces to
 \be
\dot{\xi}+ q\xi =  \dot{h},
 \ee
 \be
\dot{\beta}+ \frac{3}{10}q\beta =
-\frac{1}{10}q\xi.
 \ee
The solutions  $\beta$ and $\xi$ will be independent of the angle $\mu$.
In general case of $k \neq 0$,
the function $\beta$ and $\xi$
to be determined by Eqs.(\ref{eqn:rewritten1}) and (\ref{eqn:rewritten})
will depend on $\mu$.
As the right hand side of Eq.(\ref{eqn:rewritten})
contains an integral over $d \mu'$,
it is difficult to give an exact solution.
And one may expand $\beta$ and $\xi$ in terms of the Legendre functions
\[
\xi(\mu) =\sum_l (2l+1) \xi_l P_l(\mu),
\]
\[
\beta(\mu) = \sum_l (2l+1) \beta_l P_l(\mu),
\]
with the Legendre components
\be\label{xil}
\xi_l(\eta) = \frac{1}{2}\int_{-1}^1 \,d\mu\, \xi(\eta,\mu)P_l(\mu),
\ee
\be\label{betal}
\beta_l(\eta) =\frac{1}{2}\int_{-1}^1\, d\mu\, \beta(\eta,\mu)P_l(\mu).
\ee
The differential equations Eqs.(\ref{eqn:rewritten1}) and
(\ref{eqn:rewritten}) for $\xi(\eta,\mu)$ and
$\beta(\eta,\mu)$ then become an infinite set of coupled
differential equations for $\xi_l(\eta)$ and $\beta_l(\eta)$.

~

\begin{center}
{\em\Large 3. Decomposition of Polarization $\beta$ into Electric and Magnetic Types }
\end{center}

From Eq.(\ref{f1+}) for the definition of $\zeta$ and $\beta$  for
the $h^+$ GW polarization , it is seen that the Stokes parameters
can be be written as the following \cite{kami,cabe},
 \be
 Q(\theta,\phi)=\frac{T_0}{4}\sum_l(2l+1)P_l(\cos\theta)(1+\cos^2\theta)\cos
 2\phi ~\beta_l; \label{Q}
 \ee
 \be
 U(\theta,\phi)=\frac{T_0}{4}\sum_l(2l+1)P_l(\cos\theta)2\cos\theta\sin
 2\phi ~\beta_l. \label{U}
 \ee
One commonly uses the spherical harmonic functions $Y_{(lm)}$ as the
complete orthonormal basis for scalar functions
defined on the 2-sphere, such as the temperature anisotropies $\Delta T$.
For the $2\times 2$ tensors defined on the 2-sphere,
such as $P_{ab}$ in Eq.(\ref{tensor}),
the following complete orthonormal set of tensor
spherical harmonics can be  employed \cite{kami}:
\begin{equation}
     Y_{(lm)ab}^{\rm G} = N_l
     \left( Y_{(lm):ab} - {1\over2} g_{ab} Y_{(lm):c}{}^c \right),
\label{Yplusdefn}
\end{equation}
\begin{equation}
     Y_{(lm)ab}^{\rm C} = { N_l \over 2}
     \left(\vphantom{1\over 2}
       Y_{(lm):ac} \epsilon^c{}_b +Y_{(lm):bc} \epsilon^c{}_a \right),
\label{Ytimesdefn}
\end{equation}
where  ``$:$'' denotes
covariant derivative on the 2-sphere,
$ N_l \equiv \sqrt{ {2 (l-2)! / (l+2)!}}$, and
\begin{equation}  \label{metric}
\epsilon^a\, _b= \left(
\begin{array}{cc}
0 & \sin\theta\\
-1/\sin\theta & 0\\
\end{array}
\right). \nonumber
\end{equation}
They satisfy
\begin{equation}
      \int d\hat{n}\,Y_{(lm)ab}^{{\rm G}\,*}(\hat{n})\,Y_{(l'm')}^{{\rm
      G}\,\,ab}(\hat{n})
      =\int d\hat{n}\,Y_{(lm)ab}^{{\rm C}\,*}(\hat{n})\,Y_{(l'm')}^{{\rm
      C}\,\,ab}(\hat{n})
      =\delta_{ll'} \delta_{mm'},\nonumber
\end{equation}
\begin{equation}
     \int d\hat{n}\,Y_{(lm)ab}^{{\rm G}\, *}(\hat{n})\,
     Y_{(l'm')}^{{\rm C}\,\,ab}(\hat{n})
     =0.
\label{norms}
\end{equation}
By construction  one sees that  $ Y_{(lm)ab}^{\rm C} $ is the gradients (electric type)
and $ Y_{(lm)ab}^{\rm C} $ is the curls (magnetic type)
of the ordinary spherical harmonics, and they represent
the electric type and magnetic type components of the polarization, respectively.
The polarization tensor can be expanded in terms of this basis as:
\begin{equation}
     {{ P}_{ab}(\hat{n})\over T_0} = \sum_{l=2}^\infty\sum_{m=-l}^l
     \left[ a_{(lm)}^{\rm G}
     Y_{(lm)ab}^{\rm G}(\hat{n}) + a_{(lm)}^{\rm C} Y_{(lm)ab}^{\rm C}
     (\hat{n}) \right].
\label{abc}
\end{equation}
The expansion coefficients are given by
\begin{equation}
     a_{(lm)}^{\rm G}={1\over T_0}\int \, d\hat{n}\, {P}_{ab}(\hat{n})
              Y_{(lm)}^{{\rm G} \,ab\, *}(\hat{n}), \quad
     a_{(lm)}^{\rm C}={1\over T_0}\int d\hat{n}\, {P}_{ab}(\hat{n})
              Y_{(lm)}^{{\rm C} \, ab\, *}(\hat{n}),
\label{defmoments}
\end{equation}
and calculation yields  \cite{kami,cabe}:
\footnote{There is a small mistake in the formulae
(4.39),(4.40),(4.41) and (4.42) in Ref\cite{kami} and formulae
(111),(114) and (115) in Ref\cite{cabe}, the coefficient of
$\frac{6l(l+1)}{(2l+3)(2l-1)}$ should be replaced by
$\frac{6(l-1)(l+2)}{(2l+3)(2l-1)}.$}
\begin{eqnarray}
     a^{\rm G}_{lm} = & & \frac{1}{8}  (\delta_{m2}+\delta_{m,-2})
     \sqrt{2\pi(2l+1)} \nonumber \\
     \times & & \left[\frac{(l+2)(l+1)\beta_{l-2}}{(2l-1)(2l+1)}+\frac{6
     (l-1)(l+2) \beta_{l}}{(2l+3)(2l-1)} + \frac{l(l-1)\beta_{l+2}}
     {(2l+3)(2l+1)}\right], \label{aG}
\end{eqnarray}
\be
     a^{\rm C}_{lm}~=~\frac{-i}{4} \sqrt{\frac{2\pi}{(2l+1)}}
     (\delta_{m2}-\delta_{m,-2})[(l+2)\beta_{l-1}+(l-1)\beta_{l+1}].
     \label{aC}
\ee
where the polarization $\beta$ shows up explicitly in
the G and C components.
Then the magnetic type of power spectrum $C^{\rm GG}_l$
is \footnote{ In the Ref.\cite{kami,cabe}, the bracket of
[~] should be replaced by the absolute value sign $|~|$, which is
for the values of $\beta$ may being imaginary numbers.}
\begin{eqnarray}
     C_l^{\rm GG} &=& \frac{1}{2l+1} \sum_m\left|a^G_{lm}\right|^2 \nonumber
     \\
     &=& \frac{\pi}{16}
     \left|\frac{(l+2)(l+1)\beta_{l-2}}{(2l-1)(2l+1)}+\frac{6
     (l-1)(l+2)\beta_{l}} {(2l+3)(2l-1)}+\frac{l(l-1) \beta_{l+2}}
     {(2l+3)(2l+1)}\right|^2
\end{eqnarray}
and similarly for $C_l^{\rm CC}$.
Note that these expressions are for the gravitational wave in the
$\hat{z}$ direction with `+' polarization.
Summing over all Fourier modes,
and over both polarization states, one has the final result
\begin{equation}
     C_l^{\rm GG}=\frac{1}{16\pi}\int \,
 \left|\frac{(l+2)(l+1)\beta_{l-2}}{(2l-1)(2l+1)}+\frac{6
     (l-1)(l+2)\beta_{l}} {(2l+3)(2l-1)} +
     \frac{l(l-1)\beta_{l+2}}{(2l+3)(2l+1)}\right|^2\,k^2dk,\label{gg}
\end{equation}
 \be
 C_l^{\rm CC}=\frac{1}{4\pi}\int\,
   \left|\frac{(l+2)\beta_{l-1}}{2l+1}+\frac{(l-1)\beta_{l+1}}
  {2l+1}\right|^2\,k^2dk . \label{cc}
\ee Note that the cross-correlation power spectrum vanishes
\begin{equation}
     C_l^{\rm GC}= \sum_{m=-l}^{m=l}\frac{a^{{\rm
     G}*}_{lm}a^{\rm C}_{lm}}{2l+1}=0,
\end{equation}
since
 $a_{(lm)}^{\rm G} \propto
(\delta_{m,2}+\delta_{m,-2})$, while $a_{(lm)}^{\rm C} \propto
(\delta_{m,2}-\delta_{m,-2})$.

~

\begin{center}
{\bf\Large  4.~~Evolution of Gravitational Waves }
\end{center}

Now we consider the evolution of the relic GW,
which is the source term of the CMB
polarization in Eq.(\ref{eqn:rewritten1}).
For both polarizations,  $+, \times$,
the equation of motion for the relic GW of mode $k$ is the following:
 \be \label{h}
 \ddot{h}+2\frac{\dot{a}}{a}\dot{h}+k^2h=0,
 \ee
and the initial condition is taken to be
 \be \label{hi}
 h(\eta=0)=h(k),~~\dot{h}(\eta=0)=0,\label{ih1}
 \ee
with
 \be \label{powerh}
 \frac{k^3}{2\pi^2}  |h(k)|^2  =P_h(k)=A_T \left(\frac{k}{k_0}\right)^{n_T}, \label{ih2}
 \ee
where $P_h(k)$ is the the primordial power spectrum of GW, $A_T$
is the amplitude, $k_0= 0.05$ Mpc$^{-1}$ is the pivot wavenumber,
and $n_T$ is the the tensor spectrum index. Inflationary models
generically predicts $n_T\approx0$, a nearly scale-invariant
spectrum. Later we will also see the influence of $n_T$ on the CMB
polarization. We have ignored a suppressing effect on the
gravitational waves by the neutrinos free streaming
\cite{weinberg,abc}, which can slightly bring down the hight of
the peak at small scales \cite{prit}.

The equation (\ref{h}) depends on the scale factor $a(\eta)$,
which is determined by the  Friedmann equation
 \be\label{friedmann}
 \dot a ^2 =H_0^2 \left[\Omega_r+a\Omega_m
+a^4\Omega_{\Lambda} \right],
 \ee
where $H_0$ is the present Hubble parameter, $\Omega_r$,
$\Omega_m$, and $\Omega_\Lambda$ are the present fractional
densities for the radiation (including the photon and neutrino),
matter, and dark energy, respectively. Given a set of these
densities, one solves Eq.(\ref{friedmann}) for $a(\eta)$
numerically. Taking $\Omega_r = 8.36\times 10^{-5}$, $\Omega_m=
\Omega_b + \Omega_{dm}=0.044 +0.226$, $\Omega_{\Lambda}=0.73$, we
solve Eq.(\ref{friedmann}), and substitute the resulting $a(\eta)$
and $\dot{a}(\eta)$ into Eq.(\ref{h}), then the numerical
solutions $h(\eta)$ and $\dot{h}(\eta)$ are obtained
straightforwardly. The resulting $h(\eta_d)$ and $\dot{h}(\eta_d)$
at the decoupling time $\eta_d$ are given as function of $k$ in
Fig.1 and Fig.2, respectively.

Besides the numerical solution, we may use the simple
approximations of  $a(\eta)$ transiting between different stages
suddenly,
 \be \label{a}
 a(\eta)= \left\{
 \begin{array}{c}
 a_r\eta ,~~~~~~~~~~~~~~\eta<\eta_{e}~~~ (radiation ~ dominant),\\
 a_m\eta^{2} ,~~~~~~~~\eta_{e}<\eta\leq\eta_E~~~~ (matter ~dominant),\\
 a_l \eta^{-1}, ~~~~~~~~~~~\eta> \eta_E~~~~~~~~~~~~~ (\Lambda~ dominant)
 \end{array}
 \right.
 \ee
where $a_r$,  $a_m$, and $a_l$  are constant numbers, and can be
determined by jointing $a(\eta)$ at $\eta_{e}$ and $\eta_E$. In
the $\Lambda$CDM model\footnote{In all this paper, we choose the
present Hubble parameter $h_0=0.72$. } with $\Omega_b=0.044$,
$\Omega_{dm}=0.226$, $\Omega_{\Lambda}=0.73$, taking the redshift
$z_e=3234$ yields the conformal time $\eta_e/\eta_0=0.007$ (where
$\eta_0$ is the present conformal time), and taking $z_E=0.39$
yields $\eta_E/\eta_0=0.894$, and $\eta_d/\eta_0=0.0195$ is the
decoupling time at the redshift $z_d=1089$. Eq.(\ref{h}) has the
analytic solution \cite{grishchuk,grish,zhang}:
 \be
 h(\eta)=A_0j_0(k\eta),~~~~~(\eta<\eta_e),
 \ee
 \be
 h(\eta)=A_0(\eta_e/\eta)[A_1j_1(k\eta)+A_2y_1(k\eta)],
           ~~~(\eta_e<\eta\leq\eta_E),\label{36}
 \ee
with the coefficient
 \be
 A_0=(A_Tk^{n_T-3})^{1/2},
 \ee
determined by the primordial power spectrum, and
 \be
 A_1=\frac{3k\eta_{e}-k\eta_{e}
\cos(2k\eta_{e})+2\sin(2k\eta_{e})}{2k^2\eta_{e}^2}, \ee \be
A_2=\frac{2-2k^2\eta_{e}^2-2\cos(2k\eta_{e})-
k\eta_{e}\sin(2k\eta_{e})}{2k^2\eta_{e}^2}.
 \ee
These  $h(\eta_d)$ and $\dot{h}(\eta_d)$ are plotted with solid
lines in Fig.1 and Fig.2,
where the initial normalization $h(k)=1$ has been taken.
The figures show that the simple
approximation is good only in long wavelengths,
but it differs from the numerical one considerably
in short wavelength ($k>10$).

As an improvement to (\ref{a}), one can use the WKB approximation
for $a(\eta)$ \cite{wkb,prit}. Since we are only interested in the
gravitational waves at time $\eta_d$, and the dark energy is small
and can be omitted at that time, so the scale factor can be
approximated by
 \be \label{aa}
 a(\tau)=a_e\tau(\tau+2),
 \ee
where $\tau\equiv(\sqrt{2}-1)\eta/\eta_e$, and $a_e$ is determined
by $a_0/a_e=1+z_e$. When $\tau\ll2 $,  $a(\tau)\rightarrow \tau$,
the radiation dominated stage, and when $\tau\gg2 $,
$a\rightarrow\tau^2$, the matter dominated stage. $a(\tau)$
transits between these two stages is smooth. Then the evolution of
the gravitational waves become
 \be
 h''+2\frac{a'}{a}h'+r^2h=0,\label{hq}
 \ee
where $r\equiv k\eta_e/(\sqrt{2}-1)$,
the prime denotes $d/d\tau$.
This equation has an analytic solution discussed in Ref.\cite{prit},
which is rather complex.
In this paper, we
employ  Eq.(\ref{aa}) as an approximation of $a(\eta)$
to find the numerical solution of  Eq.({\ref{hq}}).
The resulting $h(\eta_d)$ and $\dot{h}(\eta_d)$ in
this WKB approximation are plotted with the dashed lines in  Fig.1 and Fig.2,
which show that the results are very good when comparing with the numerical ones,
and the difference of them $\ll 1\%$.
The approximation of (\ref{aa}) is simpler than the numerical $a(\eta)$,
and has better precision than the simple approximation of (\ref{a}),
and will be used to calculate the CMB polarization power spectrum.

~

\begin{center}
{\bf\Large  5.~~The Model of Decoupling History. }
\end{center}

Consider the decoupling history of the Universe.
Before the decoupling,
the  ionized baryons are tightly coupled to photons
by Thompson scattering.
Once the temperature falls below a few
eV, it becomes favorable for electrons and ions to recombine to
form neutral atoms.  As the number of charged particles falls, the
mean free path of any given photon increases. Eventually, the mean
free path becomes comparable to the horizon size and the photon
and baryon fluids are essentially decoupled,
and the CMB photons last scatter.
One can solve the ionization equations during the
recombination stage to obtain the visibility function $V(\eta)$,
which describes the probability that a given
photon last scattered from a particular time \cite{peebles, jones},
depending on the cosmological parameters, especially $\Omega_b$
and the present Hubble parameter $H_0$ \cite{Hu}.
In terms of the
optical depth $\kappa$, this visibility function is given by
\begin{equation}
\label{V}
V(\eta)=q(\eta)e^{-\kappa(\eta_0,\eta)},
\end{equation}
satisfying
\be \label{constraint}
 \int_{0}^{\eta_0}V(\eta)d\eta=1,
\ee
where the optical
depth function $\kappa(\eta_0,\eta)$ is related to
the differential optical depth $q(\eta)$
by
$q(\eta) =-d\kappa(\eta_0,\eta)/d \eta $.
Fig.3 shows the profile of $V(\eta)$ from the numerical result by cmbfast,
which  is sharply peaked around the last scattering.
In practical calculation  it is usually fitted by
a narrow Gaussian form \cite{zalda,prit}
\begin{equation} \label{v}
V(\eta)=V(\eta_d) \exp\left(-\frac{(\eta-\eta_d)^2}{2
\Delta\eta_d^2}\right),
\end{equation}
where $V(\eta_d)$ is the amplitude at the the decoupling time
$\eta_d$, and $\Delta\eta_d$ is the thickness of decoupling. The
analysis of the WMAP data \cite{map1-para} gives the redshift
thickness of the decoupling $\Delta z_d=195\pm2$, which
corresponds to $\Delta\eta_d/\eta_0=0.00143$. Then,  taking
$V(\eta_d)\eta_0=279$ and in (\ref{v}) yields a fitting shown in
Fig.3, which has large error on both side of $\eta_d$, compared
with the numerical one. To improve the fitting  of $V(\eta)$, we
take the following analytic expressions, which consists of two
half-gaussian functions,
\begin{equation}\label{halfgaussian1}
V(\eta)=V(\eta_d) \exp\left(-\frac{(\eta-\eta_d)^2}{2
\Delta\eta_{d1}^2}\right),~~~(\eta<\eta_d);
\end{equation}
\begin{equation}\label{halfgaussian2}
V(\eta)=V(\eta_d) \exp\left(-\frac{(\eta-\eta_d)^2}{2
\Delta\eta_{d2}^2}\right),~~~(\eta>\eta_d);
\end{equation}
with $\Delta\eta_{d1}/\eta_0=0.00110$,
$\Delta\eta_{d2}/\eta_0=0.00176$, and
$(\Delta\eta_{d1}+\Delta\eta_{d2})/2=\Delta\eta_{d}$,
satisfying the constraint of (\ref{constraint}).
Fig.3  shows that the half-gaussian model fits the numerical one
much better than the Gaussian fitting,
especially, on the left-hand side of the peak
the area enclosed between the curves of
the Gaussian and the half-gaussian is about $\sim 11\%$
of the total area enclosed under the curve $V(\eta)$.
As shall be seen later,
this difference in $V(\eta)$ will subsequently cause
a variation in the hight of the polarization spectrum.
The expressions (\ref{halfgaussian1}) and (\ref{halfgaussian2})
will be used to calculate
the approximate analytic  polarization power spectrum,
which turns out to depend   more sensitively  on
the smaller time interval $\Delta\eta_{d1}$.

~

\begin{center}
{\bf\Large  6.~~Analytic Solution.}
\end{center}

We start to look for the analytic solution of
Eqs.(\ref{eqn:rewritten1}) and (\ref{eqn:rewritten}). Since the
blackbody spectrum $f(\nu_0)$ in the Rayleigh-Jeans zone has the
property $\frac{d\ln f_0(\nu_0)}{d\ln\nu_0}\approx 1$, these
equations reduce to
 \be \label{xi}
 \dot{\xi}+\left[ik\mu+q\right]\xi =
 \dot{h},
 \ee
 \be \label{betae}
 \dot{\beta}+[ik\mu+ q]\beta =\frac{3q}{16} \int^1_{-1}
     d\mu'\left[(1+\mu'^2)^2\beta-\frac{1}{2}(1-\mu'^2)^2\xi\right].
 \ee
In Eq.(\ref{xi}) the  GW $\dot h$  play the role of source  for
the anisotropies $\xi$,
while the term $q\xi$ causes  $\xi$ to damp.
The formal solution of Eq.(\ref{xi}) is
\be
 \xi(\eta)=\int_0^{\eta}   \dot{h}(\eta')
 e^{-\kappa(\eta,\eta')}
e^{ik\mu(\eta'-\eta)}
 d \eta'.
 \ee
In  Eq.(\ref{betae}) the integration over $\mu$
contains the functions $\beta$ and $\xi$.
Using $\xi_l$ and $\beta_l$  defined in Eqs.(\ref{betal}) and (\ref{xil}),
then Eq.(\ref{betae}) is
\be
\dot{\beta}+[ik\mu+ q]\beta = qG,
\ee
where
\[
G(\eta) \equiv
\frac{3}{35}\beta_4 +\frac{5}{7}\beta_2 +\frac{7}{10}\beta_0
-\frac{3}{70}\xi_4 +\frac{1}{7}\xi_2-\frac{1}{10}\xi_0.
\]
One might write down a formal solution
\be\label{G}
\beta(\eta)=\int_0^{\eta}  G(\eta') q(\eta')
 e^{-\kappa(\eta,\eta')}
e^{ik\mu(\eta'-\eta)}
 d \eta',
\ee
and set the the time $\eta$ in the above to be the present time $\eta_0$,
\be\label{G1}
\beta(\eta_0) = \int_0^{\eta_0}
G(\eta')  V(\eta')  e^{ik\mu(\eta'-\eta_0)}   d \eta',
\ee
where
$V(\eta')=q(\eta')e^{-\kappa(\eta_0,\eta')}$ is the visibility function.
However, the difficulty with the integration (\ref{G1}) is that
the integrand $G$ contains $\beta_l$ and $\xi_l $ up to $l=4$,
which are not known yet.

One uses the Legendre expansion and write Eqs.(\ref{xi}) and (\ref{betae})
as the following hierarchical  set of equations:
 \be
 \dot{\xi_0}=-q\xi_0-ik\xi_1+\dot{h},\label{1}
 \ee
 \be
 \dot{\beta_0}=-\frac{3}{10}q\beta_0-ik\beta_1+q\left(\frac{3}{35}\beta_4 +\frac{5}{7}\beta_2
-\frac{3}{70}\xi_4 +\frac{1}{7}\xi_2-\frac{1}{10}\xi_0\right),
 \ee
 \be
 \dot{\xi_l}   =  -q\xi_l-\frac{ik}{2l+1}\left[l\xi_{l-1}+(l+1)\xi_{l+1}\right],
 ~~~for~~l\geq1,
 \ee
 \be
 \dot{\beta_l} =  -q\beta_l-\frac{ik}{2l+1}\left[l\beta_{l-1}+(l+1)\beta_{l+1}\right],
 ~~~for~~l \geq 1.\label{2}
 \ee

In  the tight coupling limit with $q\rightarrow\infty$,
the  equations reduce to
  \be
 \dot \xi_0+q\xi_0=\dot{h} ,
  \ee
  \be
  \dot\beta_0+\frac{3}{10}q\beta_0=-\frac{1}{10}q\xi_0,
  \ee
  \be
  \xi_l=\beta_l=0,~~~l\geq  1.
  \ee
Then the source function $G(\eta)$ reduces to
$G= (7\beta_0-\xi_0)/10$,
and satisfies the  equation:
 \be
 \dot{G}+\frac{3}{10}qG=-\frac{1}{10}\dot{h},
 \ee
and the formal solution is
\be \label{G0}
G(\eta)
=  - \frac{1}{10}   \int_0^{\eta} \dot h (\eta'')
e^{-\frac{3}{10}\kappa (\eta, \eta'')}d \eta''.
\ee
Substitute
this expression of $G$ into  Eq.(\ref{G1}), yields the formal
solution for the polarization in the tight coupling limit:
\ba \label{beta0}
\beta(\eta_0) &  =  &
\int_{0}^{\eta_0}V(\eta')
   \left[ -\frac{1}{10} \int^{\eta'}_0 \dot h(\eta'')
   e^{-\frac{3}{10}\kappa(\eta',\eta'')}d\eta''
   \right ]
      e^{ik\mu(\eta'-\eta_0)}d\eta'  \nonumber \\
& = &  -\frac{1}{10}\int_0^{\eta_0}d\eta' V(\eta')
      e^{ik\mu(\eta'-\eta_0)}
      \int_0^{\eta'} d\eta''\dot{h}(\eta'')e^{-\frac{3}{10}\kappa(\eta'')+
\frac{3}{10}\kappa(\eta')},
\ea
where $\kappa(\eta) \equiv \kappa(\eta_0,\eta)$.

Note that this approximate result of the  tight coupling
applies only on scales much larger than the mean free path of photons.
On smaller scales the effects of photon diffusion
will cause some damping in the anisotropy and polarization.
To take care of this effect,
we now expand Eqs.(\ref{1})-(\ref{2}) to the second order
of  the small parameter $1/q<<1$,
which has the meaning of the mean free path,  and arrive at,
 \be \label{diffus1}
 \dot{\xi_0}=-q\xi_0-ik  \xi_1+  \dot{h},
 \ee
 \be \label{diffus2}
  \dot{\xi_1}=-q\xi_1- \frac{ik}{3}  \xi_0,
 \ee
 \be
 \xi_l=0,~~~~l\geq 2.
 \ee
Putting $\xi_0\propto e^{i\int\omega  d\eta}$
and $\xi_1\propto e^{i\int\omega  d\eta}$ and
substituting
into Eqs.(\ref{diffus1}) and (\ref{diffus2}),
ignoring variations of $\omega$ on the
expansion scale $\dot{a}/a$,  neglecting  $\dot{h}$ which is
nearly zero at low  frequency, shown in Fig.2, one gets
 \[
 \omega=\pm\frac{k}{\sqrt{3}}+iq .
 \]
Thus $\xi_0$ will acquire an extra damping factor $e^{-\int qd \eta}$,
independent on the wavenumber $k$.
This
feature is different from the case of the scalar
perturbations, where the damping is strong on the small scales \cite{hu1}.
For the polarization $\beta$, we only keep
the tight coupling limit with the equation,
 \[
\dot{\beta_0}=-\frac{3q}{10}\beta_0 -\frac{q}{10}\xi_0,
 \]
% \[
% \frac{1}{q} \dot{\beta_1}= -\beta_1 -\frac{ik}{3} \frac{1}{q} \beta_0,
% \]
% \[
%\beta_l=0,~~~~l\geq 2.
% \]
which follows that $\beta_0$ also gets a damping of $e^{-\int qd \eta}$.
Thus, taking into the account of the diffusion effect,
$G$ in (\ref{G0}) will acquire an extra damping factor $\exp(-\kappa(\eta))$,
and  (\ref{beta0}) becomes
 \be
 \beta(\eta_0)   =   -\frac{1}{10}\int_0^{\eta_0}d\eta' V(\eta')
 e^{ik\mu(\eta'-\eta_0)}\int_0^{\eta'}
 d\eta''\dot{h}(\eta'')e^{-\frac{3}{10}\kappa(\eta'')-
 \frac{7}{10}\kappa(\eta')}.
 \ee

In the above
the function $\exp{(-\frac{3}{10}\kappa(\eta''))}\simeq 0$ for
$\eta<\eta_d$,
and
$\exp{(-\frac{3}{10}\kappa(\eta''))}\simeq 1$ for $\eta>\eta_d$,
so it can be approximated as a step function
$\exp{(-\frac{3}{10}\kappa(\eta''))}  \approx \theta(\eta''-\eta_d )$,
and moreover,
the visibility function
$V(\eta')$ is also peaked about the decoupling $\eta_d$.
Therefore,  as an approximation,
one can pull the $\dot h(\eta'')$ out of the integration $\int d\eta''$,
 \be
 \beta(\eta_0)=  -\frac{1}{10}\int_0^{\eta_0}d\eta V(\eta)
e^{ik\mu(\eta-\eta_0)}\dot{h}(\eta)\int_0^{\eta}d\eta'e^{-\frac{3}{10}\kappa(\eta')-
\frac{7}{10}\kappa(\eta)}.
 \ee
Define the integration variable
$x \equiv \kappa(\eta')/\kappa(\eta)$ to replace the variable $\eta'$.
For $V(\eta)$ is peaked  around the $\eta_d$ with width $\Delta\eta_d$,
one can take $d\eta'=\frac{dx}{x}\Delta\eta_d$ as an  approximation,
then
 \be\label{65}
 \beta(\eta_0)=-\frac{1}{10}\Delta\eta_d  \int_0^{\eta_0}d\eta V(\eta)
e^{ik\mu(\eta-\eta_0)}\dot{h}(\eta)\int_1^{\infty}\frac{dx}{x}e^{-\frac{3}{10}\kappa(\eta)
x}e^{-\frac{7}{10}\kappa(\eta)},
 \ee
Substituting
this into Eq.(\ref{betal})and using the  expansion formula
\[
 e^{ikr\cos\theta}
      =\sum_{l=0 }^{\infty} (2l+1)i^lj_l(kr)P_l(\cos\theta),
\]
one arrives at the expression for the component of the polarization
\be \label{bl}
 \beta_l(\eta_0)=-\frac{1}{10}  \Delta\eta_d\, i^l
 \int_0^{\eta_0}d\eta V(\eta) \dot{h}(\eta)   j_l(k(\eta-\eta_0))
 \int_1^{\infty}\frac{dx}{x}e^{-\frac{3}{10}\kappa(\eta)
x}e^{-\frac{7}{10}\kappa(\eta)}.
\ee

Now look at the integration $\int d\eta$ involving
$V(\eta)$, which has a factor of the form $e^{-a(\eta-\eta_d)^2}$.
As a stochastic quantity,  the time-derivative of GW $\dot{h}(\eta)$
contains generally a mixture of oscillating modes,
such as
$e^{ik\eta}$ and $e^{-ik\eta}$,
and so does the spherical
Bessel function $j_l(k(\eta-\eta_0))$.
Thus $\dot{h}(\eta)   j_l(k(\eta-\eta_0)) $ generally contains
terms $\propto e^{-ibk(\eta-\eta_0)}$, where $b\in [-2, 2]$.
Using the formula
%\[
%e^{-a\eta^2+ibk\eta} = e^{-\frac{(bk)^2}{4a}}
%e^{-a(\eta-i\frac{bk}{2a})^2 },
%\]
\[
\int_{-\infty}^{\infty}
e^{-ay^2}e^{ibky}dy=e^{-\frac{(bk)^2}{4a}}\int_{-\infty}^{\infty}
e^{-ay^2}dy,
\]
the integration is approximately
 \be
 \int_0^{\eta_0}d\eta V(\eta) \dot{h}(\eta)   j_l(k(\eta-\eta_0))
 \approx
e^{-\alpha(k\Delta\eta_d)^2}\dot{h}(\eta_d) j_l(k(\eta_d-\eta_0))
\int_0^{\eta_0}d\eta V(\eta).
 \ee
When the half-gaussian visibility function $V(\eta)$ of
Eq.(\ref{halfgaussian1}) and (\ref{halfgaussian2}) is used,
the integration  is approximated by
 \be
 \int_0^{\eta_0}d\eta V(\eta)
\dot{h}(\eta) j_l(k(\eta-\eta_0)) \approx
\frac{1}{2}\left[e^{-\alpha(k\Delta\eta_{d1})^2}+e^{-\alpha(k\Delta\eta_{d2})^2}\right]
\dot{h}(\eta_d)j_l(k(\eta_d-\eta_0)) \int_0^{\eta_0}d\eta V(\eta).
 \ee
In the above $\alpha$ can take values in $[0,2]$,
depending on the phase of
$\dot{h}(\eta)   j_l(k(\eta-\eta_0)) $.
Here we will take $\alpha$ as a parameter.
This gives another damping depending on both $\Delta\eta_d$ and $k$.
This damping of the anisotropies  $\xi$ and $\beta$ of CMB is caused by
Thompson scattering of photons by free electrons.
During the recombination around the last scattering, the visibility
function $V$ is narrowly centered around the the time $\eta_d$
with a width $\Delta\eta_d$,
so the smoothing by Thompson
scattering is effectively limited within the interval $\Delta \eta_d$.
Thus a wave of anisotropies $e^{ik\eta}$ will be damped in this interval by
an factor $e^{-(k\Delta \eta_d)^2}$.
The longer the interval $\Delta\eta_d$ is, the more damping the wave suffers.
In fact  $\Delta\eta_d$ can be viewed as the thickness of the last
scattering surface.
Those waves with wavelength $\lambda$ shorter than $\Delta\eta_d$
will be effectively damped by a factor $e^{-(2\pi \Delta \eta_d /\lambda)^2}$,
the same as before.
Thus the shorter the wave length is,
the more damping for the wave.

The remaining integrations in $\beta_l$ is
\be
 \int_0^{\eta_0}d\eta V(\eta)
 \int_1^{\infty}\frac{dx}{x}e^{-\frac{3}{10}\kappa(\eta)
  x}e^{\frac{-7}{10}\kappa(\eta)} =
  \int_0^{\infty}d\kappa e^{-\frac{17}{10}\kappa}
 \int_1^{\infty}\frac{dx}{x}e^{-\frac{3}{10}\kappa x}
=  \frac{10}{17}\ln\frac{20}{3}. \ee We like to point out that
this amplitude is the outcome from the second order of the
tight-coupling limit, while the first order of the tight-coupling limit
with $G$ being given in (\ref{G0})
would yield a result $ \frac{10}{7}\ln\frac{10}{3}$ in Ref.\cite{prit}.
Thus finally (\ref{bl}) yields
 \be \label{b}
 \beta_l(\eta_0)=
 - \frac{1}{17}\ln\frac{20}{3} i^l
 \Delta\eta_d\dot{h}(\eta_d)j_l(k(\eta_d-\eta_0))D(k),
 \ee
where \be \label{D}
D(k)\equiv\frac{1}{2}\left[e^{-\alpha(k\Delta\eta_{d1})^2}
+e^{-\alpha(k\Delta\eta_{d2})^2}\right]. \ee
for the half-gaussian
visibility function. For the gaussian  visibility function one
would have $D(k)\equiv e^{-\alpha(k\Delta\eta_d)^2}$.

Substituting this back into Eqs.(\ref{gg}) and (\ref{cc}) yields
the polarization spectra
 \be\label{power}
 C^{XX}_{l}=\frac{1}{16\pi}\left(\frac{1}{17}\ln\frac{20}{3}\right)^2
 \int  P_{Xl}^2(k(\eta_d-\eta_0))|\dot{h}(\eta_d)|^2
\Delta\eta_{d}^2 D^2(k)\,k^2dk,
\ee
where "X" denotes "G" or "C",
the type of of the CMB polarization,
for the electric type
 \be \label{pg}
 P_{Gl}(x)=\frac{(l+2)(l+1)}{(2l-1)(2l+1)}j_{l-2}(x)
 -\frac{6(l-1)(l+2)}{(2l-1)(2l+3)}j_{l}(x)+\frac{l(l-1)}{(2l+3)(2l+1)}j_{l+2}(x),
 \ee
 and for the magnetic type
 \be \label{pc}
 P_{Cl}(x)=\frac{2(l+2)}{2l+1}j_{l-1}(x)-\frac{2(l-2)}{2l+1}j_{l+1}(x).
 \ee

The result (\ref{power}) is similar to the result in
Ref.\cite{prit} (realizing that our $(2\pi)^3$ Fourier conventions
differ from theirs) if we identify $C_l^{GG}=C_{El}/2$ and
$C_l^{CC}=C_{Bl}/2$, but here the coefficient is
$\frac{1}{17}\ln\frac{20}{3}$, smaller than that in
Ref.\cite{prit} since we have also included the diffusion effect
on the source $G$. So the the second order of the tight-coupling
limit reduces the amplitude by about $\sim 58\%$. Another
difference is the damping factor $D(k)$. Besides, the parameter
$\alpha$ is taken to be in the range [0,~2], instead of a fixed
$\alpha=1/2$ as in Ref.\cite{prit}.

To completely determine $ C^{XX}_{l}$, we need to fix the
normalization  of the initial amplitude $\dot{h}(\eta_d)$ in
Eq.(\ref{power}). What has been observed is the CMB  temperature
anisotropies, which generally has contributions from both the
scalar and tensor perturbations. The ratio of the contributions
\be \label{r} r=\frac{P_h(k_0)}{P_R(k_0)} \ee has not been fixed
observationally, and only constraints can be given. Based  upon
the observations of Ly-$\alpha$ forest power spectrum from the
SDSS, 3-year WMAP, supernovae and galaxy clustering, one can give
a constraint of $r< 0.22$ at $95\%$ C.L. ($<0.37$, at $99.9\%$
C.L.) \cite{seljak}. We take it as a parameter. WMAP observation
\cite{map1-inflation} gives the scalar perturbation power spectrum
\be P_R(k_0)=2.95\times10^{-9}A(k_0), \ee with
$k_0=0.05$Mpc$^{-1}$ and the amplitude $A(k_0)=0.8$. Taking the
scale-invariant spectrum with $n_T= 0$ in (\ref{powerh}), then the
amplitude $A_T$ in (\ref{powerh}) depends on $r$. For instance, if
$r=1$ is taken, then $A_T =2.46\times 10^{-9}$, and smaller $r$
will yield smaller $A_T$ accordingly.

~

\begin{center}
{\bf\Large  7. Results and Discussions}
\end{center}

\emph{Damping Effects due to Visibility Function:}

The power spectra of $C_{l}^{GG}$ and $C_{l}^{CC}$, calculated
from our analytic formulae (\ref{power}) and from the numerical
cmbfast, have been shown in Figs.4 and 5, respectively.
The approximate analytic  result
is quite close to the numerical one evaluated from the cmbfast code,
especially, for the first three peaks of the spectrum that are observable.
One
sees that $C_{l}^{GG}$ and $C_{l}^{CC}$ at large $l$ sensitively
depend on the visibility function $V$, that is, on the factor
$D(k)$. In particular for the electric polarization spectrum,  our
half-gaussian model $D(k)$ in (\ref{D}) with
 $\alpha= 1.7$ gives very good fitting
with the third peak being very close to the numerical one, and
the second peak being higher.
For $\alpha=2$ the second peaks of the spectra are good,
but the third peaks are a bit too low.
On the other hand, the Gaussian model
$D(k)=e^{-\alpha(k\Delta\eta_d)^2}$ with $\alpha=2$
yields a power spectrum too low.
The reason is for $\Delta\eta_{d}>\Delta\eta_{d1}$.
The larger value of $\alpha$,
the larger damping of the power spectrum on the small scale.

~

\emph{The height of the power spectrums:}

 Eq. (\ref{power}) shows that the height of
 $C_l^{XX}$ depends on the amplitude
$|\dot{h}(\eta_d)|$ at the decoupling time.
As has been discussed earlier, for the
scale-invariant power spectrum  $n_T\simeq0$,
$|\dot{h}(\eta_d)|$ is directly related to
 the tensor-scalar ratio $r$ in Eq.(\ref{r}).
A larger $r$ yields a larger $|\dot{h}(\eta_d)|$  and a higher
polarization. Currently the observations have only given an upper
limit of $r<0.22$ ($95\%$ C.L.) \cite{seljak}.
In Fig.6 we have
plotted the analytic formula (\ref{power}) for $C_l^{CC}$, for
three values $r=0.3$, $0.1$, $0.01$, respectively, whereby  also
plotted are the one-sigma sensitivity estimates of the near-term
projects, WMAP and Planck satellites \cite{wp,wmap,planck}. The
WMAP estimates are based on measured noise properties of the
instrument for an $8$-year of operation, and the Planck estimates
are based on noise measurements from the test-bed High Frequency
Instrument for a $1.2$-year of operation. Fig.6 clearly shows
that the magnetic polarization for the models with $r>0.1$ can be
detected by the Planck, but difficult for the  WMAP.

However, in this discussion, we have not considered the effect of
cosmic reionization process, which is directly related to the
galaxy formation. The recent WMAP result \cite{map3-over} tends to
give the optical depth of reionization $\kappa_r=0.09\pm0.03$.
Thus the visibility function $V(\eta)$ will have another peak
around the late time $\eta/\eta_0\sim0.27$ besides the  cosmic
decoupling, and will give an extra contribution to $C_l^{CC}$,
correspondingly. At present, the reionization process is not well
understood yet, and  it is difficult to give an analytic formula
for this process. Using the numerical cmbfast including the
reionization effect, we have plotted $C_l^{CC}$  in Fig.7,
where
an extra peak of $C_l^{CC}$ at $l\sim 6$ is seen. On the
observational side, a number of other projects are currently
underway, such as CBI\cite{CBI}, DASI\cite{DASI},
CAPMAP\cite{CAPMAP}, BOOMERANG\cite{BOOM}, emphasizing  on  the
CMB  magnetic polarization. The future projects
CLOVER\cite{clover} and QUIET\cite{quiet} are expected to detect
the magnetic polarization for $r>0.01$, and the project CMBPOL
\cite{pol_net}even for $r>10^{-3}$.

~

\emph{ The influence of width of decoupling:}

Besides the gravitational waves, $C_l^{XX}$ also directly depend
on the thickness of the decoupling $\Delta\eta_d$, and on the
damping function $D(k)$.
A smaller $\Delta\eta_d$ makes the power
spectrum having smaller height. This effect is obvious at the
large scale (small $l$).
At
the small scale (large $l$), the effect is complicate
since both $\Delta\eta_d$ and the damping factor $D(k)$ will
influence the spectrum.
The expression of $D(k)$ shows
that, for a fixed $k$, the smaller $\Delta\eta_d$ leads to larger
$D(k)$.
The thickness
of the decoupling is mainly determined by the baryon density $\Omega_b$ of the
Universe.
In the flat $\Lambda$CDM Universe, increasing $\Omega_b$
will slightly enhance the decoupling speed, which will make
$\Delta\eta_d$ becoming smaller \cite{jones}.
For example, a fitting formula can be used for the optical depth function
in the $\Lambda$CDM Universe \cite{Hu}
 \be   \label{kappa}
 \kappa(z_0,z)=\Omega_b^{c_1}(\frac{z}{1000})^{c_2},~~800<z<1200,
 \ee
where $c_1=0.43$ and $c_2=16+1.8\ln\Omega_b$.
This function is
only dependent on $\Omega_b$. The
visibility function $V(\eta)=\frac{d\kappa}{d\eta}e^{-\kappa}$ is
peaked around at $\eta_d$.
A larger $\Omega_b$ corresponds to
a narrower $V$ and smaller $\Delta\eta_d$, as shown in Fig.8,
where three models
$\Omega_b=0.02,0.044,0.09$ are given.
Therefore,  a higher $\Omega_b$ leads to a lower  $C_{l}^{CC}$,
as shown in Fig.9
for these three values of $\Omega_b$.

~

\emph{The location of the peaks:}

From the formula (\ref{power}), we also can analyze the peak's
location of the power spectrums. The factor functions
$P_{Gl}^2[k(\eta_d-\eta_0)]$ in (\ref{pg}) and
$P_{Cl}^2[k(\eta_d-\eta_0)]$ in (\ref{pc}) are all combination of
the spherical Bessel function $j_l(k(\eta_d-\eta_0))$, which is
peaked at $l\simeq k(\eta_0-\eta_d)\simeq k\eta_0$ for $l\gg1$.
In
Fig.10 and Fig.11  $P_{Gl}$ and $P_{Cl}$ are plotted with $l=100$,
where it is shown  that $P_{Gl}$ peaks at $k\eta_0\simeq l$, and
$P_{Cl}$ peaks at $k\eta_0\simeq 1.27l$.
So the peak location of
the power spectrums are directly determined by
 \be \label{c}
 C_l^{XX}\propto  \left|\dot{h}(\eta_d)\right| ^2k^2D^2(k)\left.\right|_{k=l/\eta_0},\label{lg}
 \ee
The factor $D(k)$ has a larger damping at larger $l$,
so the first peak of the
power spectrum has the highest amplitude.
Let us look at the first peak of $ C_l^{XX}$,  where $D(k)\simeq1$.
Eq.(\ref{36}) gives
$\dot{h}(\eta_d)^2=A_0^2k^2(\eta_e/\eta_d)^2[A_1j_2(k\eta_d)+A_2y_2(k\eta_d)]^2$.
Since $j_2(k\eta_d)$ term is the increasing mode and the
$y_2(k\eta_d)$ term is the damping mode for the waves inside the
horizon,
so  we can take $\dot{h}(\eta_d)^2\sim
A_0^2k^2(\eta_e/\eta_d)^2[A_1j_2(k\eta_d)]^2$.
As $j_2(x)$ peaks at nearly $x\simeq3$, so $\dot{h}(\eta_d)^2$ peaks at $k\eta_d\simeq3$.
 Thus $C_l^{XX}$ peaks around
 \be \label{l}
l\simeq k\eta_0\simeq3\eta_0/\eta_d.
 \ee
This sudden-change approximation is a reasonably good estimate.
If we use
the WKB approximation in the
$\Lambda$CDM Universe with $\eta_d/\eta_0=0.0195$,
then we find that the factor function
$\dot{h}(\eta_d)^2k^2$ in Eq.(\ref{c}) is peaked around
$k\eta_0\sim 127$, shown in Fig.12,
while the
sudden-change approximation in Eq.(\ref{l}) gives $k\eta_0\sim 154$.
Therefore, the estimate (\ref{l}) holds approximately.
The value of $\eta_0/\eta_d$ is
related to the dark energy component and baryon component.
For
instance, we take  the three models  with $\Omega_{\Lambda}=0.65$,
$0.73$, and $0.80$, respectively, and with fixed $\Omega_b=0.044$,
$\Omega_{dm}=1-\Omega_{\Lambda}-\Omega_b$.
 Then a numerical
calculation yields that $\eta_0/\eta_d\simeq50.1$, $51.3$, and
$53.6$, respectively.
The gravitational waves
$\dot{h}(\eta_d)$ also depends on $\Omega_{\Lambda}$,
 as shown in Fig.13,
where it is clearly shown that a smaller $\Omega_{\Lambda}$ will shift the
peaks  $\dot{h}(\eta_d)$  slightly to
larger scales.
Correspondingly, a smaller $\Omega_{\Lambda}$ will shift the
peak of  $C_l^{XX}$ to larger scales, as demonstrated in Fig.14.
So we
have the conclusion that a lower dark energy $\Omega_{\Lambda}$
makes the peak of $C_l^{XX}$ to locate at smaller $l$.
This suggests a new
way to study the cosmic dark energy.

The baryon component also influence the decoupling time $\eta_d$.
A larger $\Omega_b$ has a larger decoupling time $\eta_d$, and
therefore, a smaller $l\simeq 3\eta_0/\eta_d$. For fixed
$\Omega_{\Lambda}=0.73$ and
$\Omega_{dm}=1-\Omega_{\Lambda}-\Omega_b$, the three models with
$\Omega_b=0.02$, $0.044$, and $ 0.09$, respectively, are given in
Fig.8. The corresponding values are $\eta_0/\eta_d=54.9$,
$51.3$,  and $ 50.1$, respectively.
So we have the conclusion that
a higher baryon  density $\Omega_b$ makes the peak to locate at
smaller $l$, as  is demonstrated  in Fig.9.

~

\emph{The influence of the  spectrum index $n_T$ of
relic GW on  $C_l^{XX}$:}

The initial condition in (\ref{hi}) of the the relic GW will influence
the CMB polarization.
Not only the value of initial amplitude $h(0)$
but also
the index $n_T$ in (\ref{powerh})
will determine $C_l^{XX}$.
To reveal how the index $n_T$ changes the $C_l^{CC}$,
we plot in Fig.15 three curves of $C_l^{CC}$
for  $n_T=-0.1$, $0.0$,
and $0.1$, respectively,
where the  parameters are taken:
$r=1$, $\alpha =2$ in the half-gaussian model.
Fig.15 shows clearly that a larger value of $n_T$ produces
a higher polarization spectrum $C_l^{CC}$.
The reason for this feature is the following:
  $C_l^{CC} \propto |\dot h(\eta_d)|^2$, and
$|\dot h(\eta_d)|$ differs from $0$ only at larger $k\eta_0 >50$,
as in Fig.2,
and $|\dot h(\eta_d)| \propto  k|h(\eta_d)| $.
Since $k^3 |h(\eta_d)|^2\propto  k^{n_T}$,
so  in the range of larger $k$ a larger value $n_T$ will
give a larger $|h(\eta_d)|^2$ and a  larger $C_l^{CC}$.
Similar behavior also occurs for the electric polarization $C_l^{GG}$.

~

\begin{center}
{\bf\Large  8. Conclusion.}
\end{center}

In this paper we have  analytically calculated
the polarization power spectrum of CMB generated by relic
gravitational waves in the Polnarev's method.
As an approximate analytic  result,
it is quite close to the numerical one evaluated from the cmbfast code,
especially, for the first three peaks of the spectrum that are observable.
We have arrived at the analytic polarization spectra $C_l^{XX}$
with several improvements over the previous results.

1. The spectrum $C_l^{XX}$
is proportional to  the relic
gravitational waves $|\dot h(\eta_d)|^2$,
the value of which is taken from  the WKB approximation
in our treatment,
which is good compared with the numerical results.
When looking for the location of the peaks of $C_l^{XX}$ analytically,
we find that it is very convenient to employ
the analytic exact solution $|\dot h(\eta_d)|^2$ from
the sudden-change approximation.
For instance, the first peak is found to be located
at $l\simeq 3\eta_0/\eta_d.$
Moreover, it has the merit that
the dependence on the dark energy $\Omega_{\Lambda}$
and the baryon $\Omega_{b}$, through the decoupling time $\eta_d$,
 is also clearly demonstrated.
Both lower dark energy $\Omega_{\Lambda}$ and
higher baryon  density $\Omega_b$
make the peak of $C_l^{XX}$ to locate at smaller $l$.

2. The visibility function describing the decoupling process
has been given by the analytic half-gaussian approximation,
which fits analytically
the actual decoupling process better than the usual Gaussian
model. For example, the improvement of $V(\eta)$ on the left side
of $\eta_d$ in the half-gaussian model is considerable, which is
about $\sim 11.5 \%$, and correspondingly, this also leads an
improvement on the spectrum $\Delta C_l^{XX}/C_l^{XX} \sim 30\%$
around the second and the third peaks.
The time integration $V(\eta)$ yields  the damping factor
$D(k)$, which contains  a parameter $\alpha$  in the range [0,~2],
coming from the phase of  the product of the $\dot{h}(\eta)$ and
$j_{l}(k(\eta-\eta_0))$.
In particular,  our half-gaussian model
 with  $\alpha \in (1.7, 2.0)$ gives a reasonably good fitting
 to the first three peaks of the spectra $C_l^{XX}$ .

3. In dealing with  the
Boltzmann equations analytically, we have worked up to the second
order of the tight coupling limit, resulting in an amplitude of
$C_l^{XX}$ smaller than that in the usual first order by $58\%$.

 5.  We have studied the influence of the
power spectrum index of the relic GW on
the CMB polarization,
and have found that
a larger value of $n_T$ produces
higher polarization spectra $C_l^{CC}$ and $C_l^{GG}$.

 ~

ACKNOWLEDGMENT: We acknowledge the using of CMBFAST
program \cite{cmbfast}.
We thank J.R.Pritchard for helpful discussions.
Y. Zhang's research work has been supported by the Chinese NSF (10173008),
NKBRSF (G19990754), and by SRFDP.

%\newpage

\newpage

\begin{figure}
\centerline{\includegraphics[width=12cm]{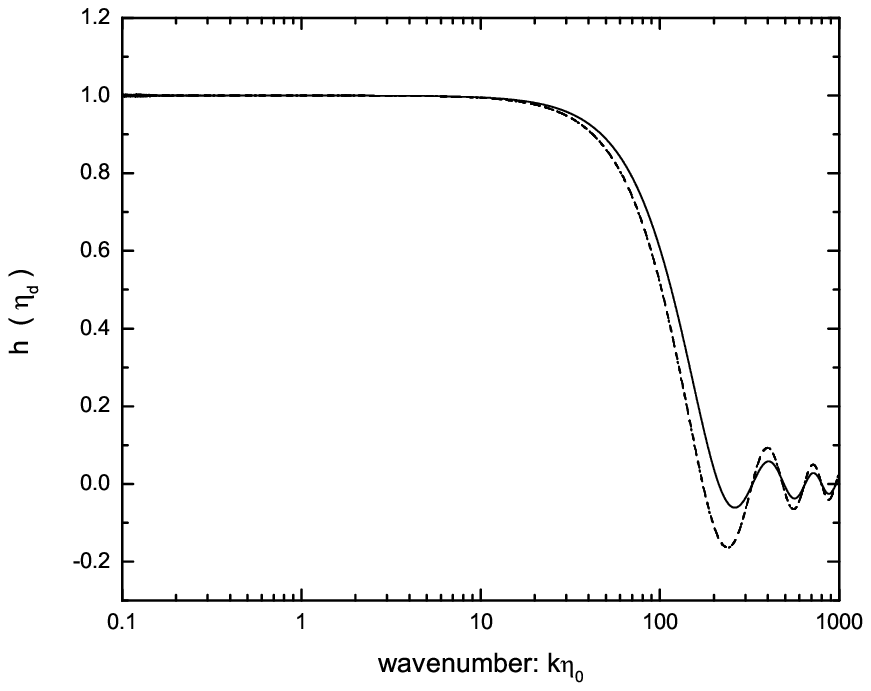}} \caption{\small
The gravitational waves $h(\eta_d)$ depends on the scale factor
$a(\eta)$. The solid line is the result of the sudden-change
approximation, the dash line is the WKB approximation result, and
the dot line is the numerical results, which is nearly overlapped
with the dash line. Note that, here the initial gravitational
waves have been rescaled with $h(k)=1$ for demonstrational
purpose. }
\end{figure}

\begin{figure}
\centerline{\includegraphics[width=12cm]{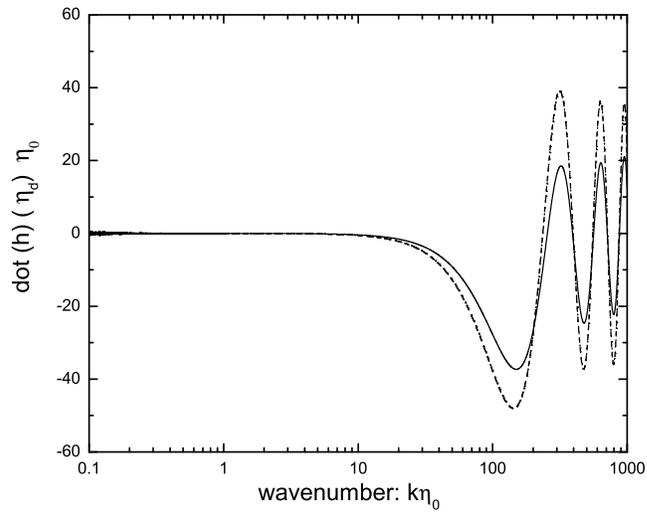}} \caption{\small
The gravitational waves $\dot h(\eta_d)$ depends on the scale
factor $a(\eta)$. The solid line is the result of the
sudden-change approximation, the dash line is the WKB
approximation result, and the dot line is the numerical results,
which is nearly overlapped with the dash line. Again the rescaled
$h(k)=1$ is used as in Fig.1. }
\end{figure}

\begin{figure}
\centerline{\includegraphics[width=12cm]{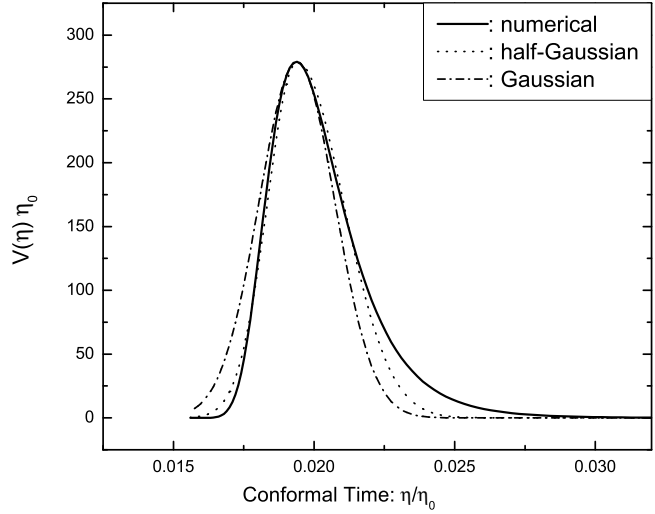}} \caption{\small
The visibility function $V(\eta)$ around the decoupling. The solid
line denotes the numerical result, the dash dot line denotes the
gaussian fitting in Eq.(\ref{v}) with
$\Delta\eta_d/\eta_0=0.00143$ and $V(\eta_d)\eta_0=279$, and the
dot line denotes our half-gaussian fitting in
Eqs.(\ref{halfgaussian1}) and (\ref{halfgaussian2}) with
$\Delta\eta_{d1}/\eta_0=0.00110$ and
$\Delta\eta_{d2}/\eta_0=0.00176$. Here $\eta_d/\eta_0=0.020$.
There is an area of strip on the left side of $\eta_d$ in the
Gaussian model differing from the numerical result, which is about
$\sim 11.5 \%$. This variation will lead to an error in the
spectrum $C_l^{XX}$ correspondingly. }
\end{figure}

\begin{figure}
\centerline{\includegraphics[width=12cm]{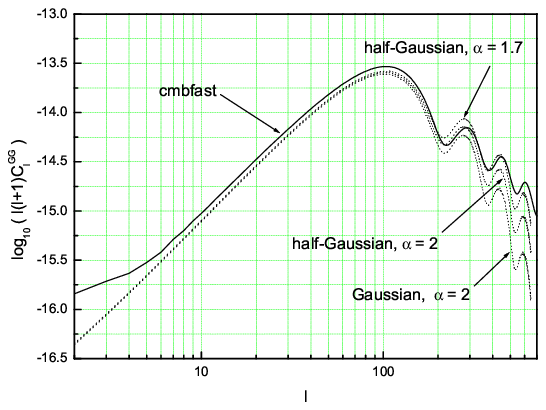}} \caption{\small
The ``electric" polarization power spectrum $C^{GG}_l$ with the
ratio $r=1$. The solid line denotes the numerical spectrum from
CMBFAST code, and the dot lines denote the analytic results. The
upper dot line is the result of the half-gaussian visibility
function with $\alpha=1.7$, the middle dot line is the result of
the half-gaussian visibility function with $\alpha=2$ and, and the
lower dot line is the result from the Gaussian visibility function
with $\alpha=2$. While at large scales these models are close to
each other, the half-gaussian model improves the spectrum height
of the $3^{rd}$  peak by about $\Delta \log_{10}C_l^{GG}\sim 0.2$
.}
\end{figure}

\begin{figure}
\centerline{\includegraphics[width=12cm]{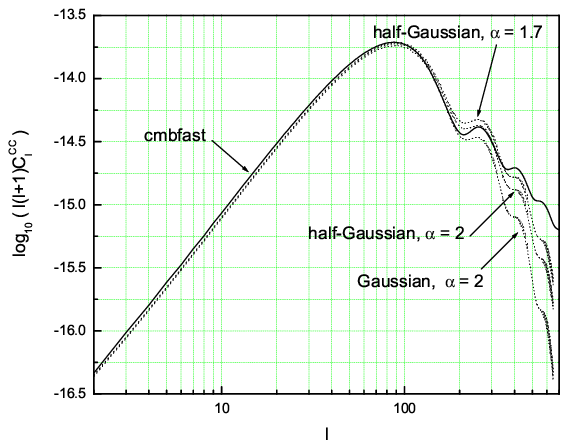}} \caption{\small
The ``magnetic'' polarization power spectrum $C^{CC}_l$ with the
ratio $r=1$. The solid line denotes the numerical spectrum from
CMBFAST code, and the dot lines denote the analytic results. The
upper dot line is the result from the half-gaussian visibility
function with $\alpha=1.7$, the middle dot line is the result from
the half-gaussian visibility function with $\alpha=2$, and the
lower one is the result from the  Gaussian visibility function
with $\alpha=2$. Again at large scales these models are close to
each other, but the half-gaussian model improves the spectrum
height of the $2^{nd}$  peak by about $\Delta
\log_{10}C_l^{CC}\sim 0.1$ .}
\end{figure}

\begin{figure} \centerline{\includegraphics[width=12cm]{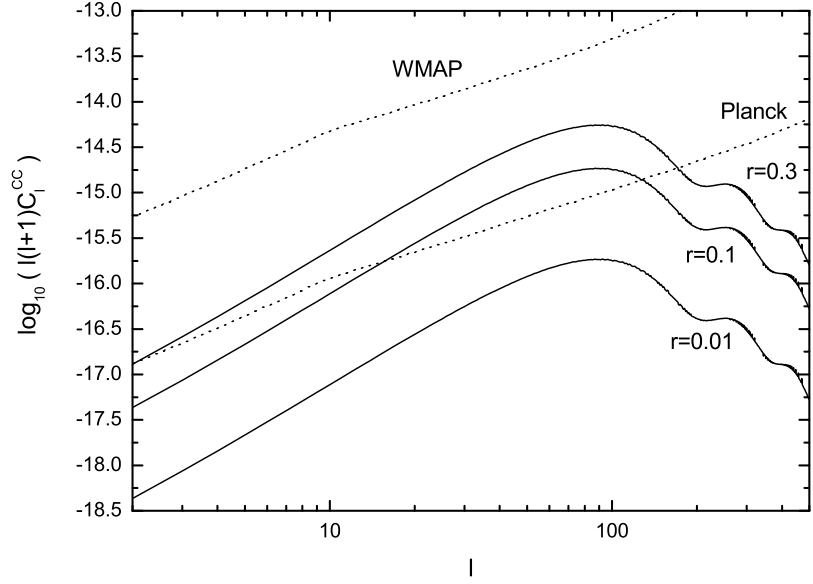}}
\caption{\small The WMAP and Planck satellite measurements on the
CMB magnetic polarization signal.
The three solid curves show the analytic
polarization power spectrums $C^{CC}_l$ for the
tensor-scalar ratio $r=0.3,0.1,0.01$, respectively,
in the $\Lambda$CDM Universe with $\Omega_b=0.044$, $\Omega_{dm}=0.226$,
$\Omega_{\Lambda}=0.73$ .  }
\end{figure}

\begin{figure} \centerline{\includegraphics[width=12cm]{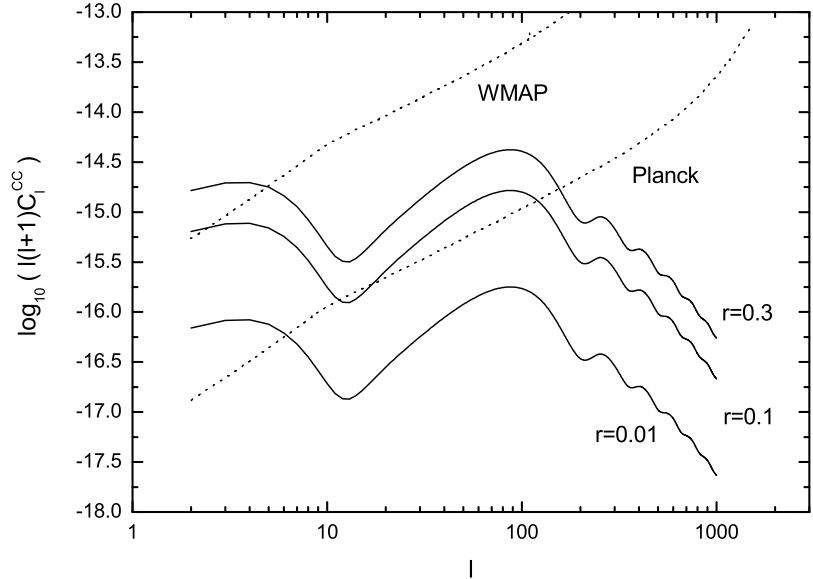}}
\caption{\small The WMAP and Planck satellite measurements on the
CMB magnetic polarization signal.
The parameters are the same as in Fig.6.
The solid lines show the numerical  $C^{CC}_l$ using cmbfast.
Here we have included the influence of
cosmic reionization with the reionization optical depth
$\kappa_r=0.09$. }
\end{figure}

\begin{figure}
\centerline{\includegraphics[width=12cm]{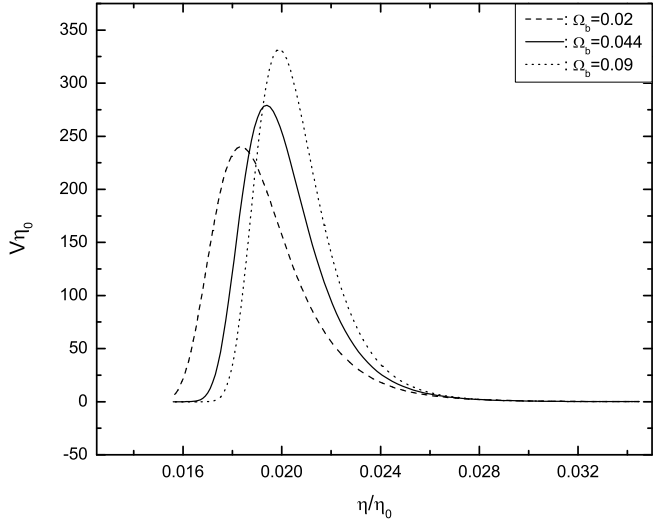}} \caption{\small
The dependence of the visibility function $V(\eta)$ on the baryon
$\Omega_b$ in $\Lambda$CDM Universe with $\Omega_{\Lambda}=0.73$,
and $\Omega_{dm}=1-\Omega_{\Lambda}-\Omega_b$. The baryon density
has been taken to be $\Omega_b=0.02$, $0.044$, $0.09$,
respectively. }
\end{figure}

\begin{figure}
\centerline{\includegraphics[width=12cm]{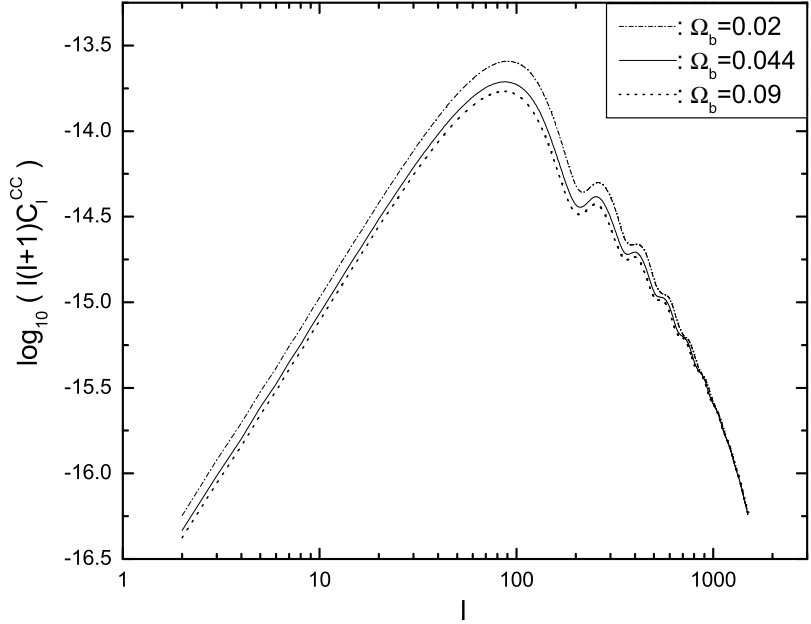}} \caption{\small
The dependence of the magnetic polarization power spectrums on the
baryon $\Omega_b$ in $\Lambda$CDM Universe with
$\Omega_{\Lambda}=0.73$,
$\Omega_{dm}=1-\Omega_{\Lambda}-\Omega_b$, and $r=1$. The baryon
$\Omega_b=0.02$, $0.044$, $0.09$, has been taken respectively, as
in Fig.8. }
\end{figure}

\begin{figure}
\centerline{\includegraphics[width=12cm]{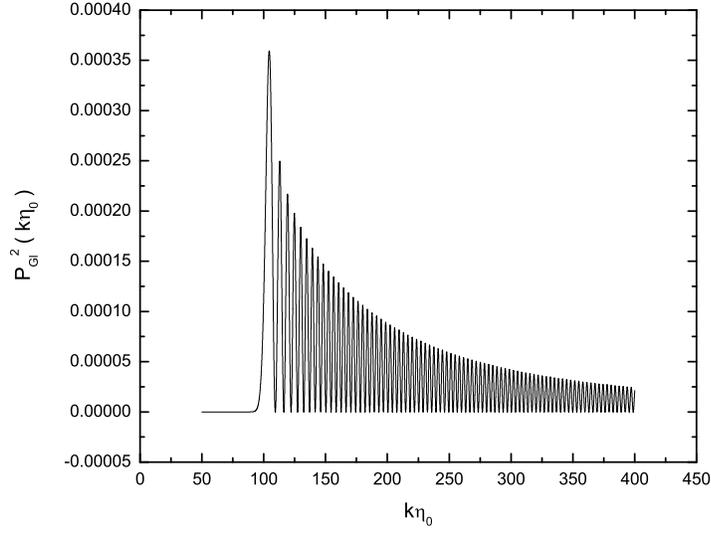}} \caption{\small
The factor  $P^2_{Gl}(k)$ as given in Eq.(\ref{pg}) with fixed
$l=100$ as a function of the wave number $k$. Obviously it is
peaked around $k\eta_0 \sim 100$, verifying the relation
(\ref{l}).}
\end{figure}

\begin{figure}
\centerline{\includegraphics[width=12cm]{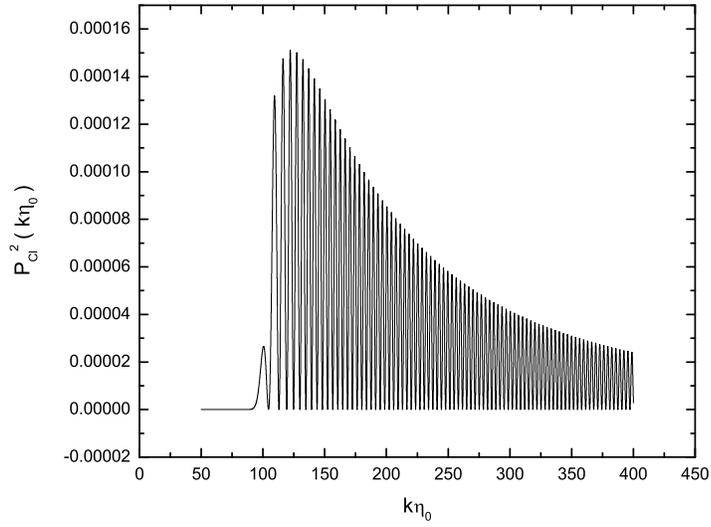}} \caption{\small
The factor  $P^2_{Cl}(k)$ as given in Eq.(\ref{pc}) with fixed
$l=100$ as a function of the wave number $k$. It is approximately
peaked around $k\eta_0 \sim 127$, thus (\ref{l}) as an estimate
holds approximately.}
\end{figure}

\begin{figure}
\centerline{\includegraphics[width=12cm]{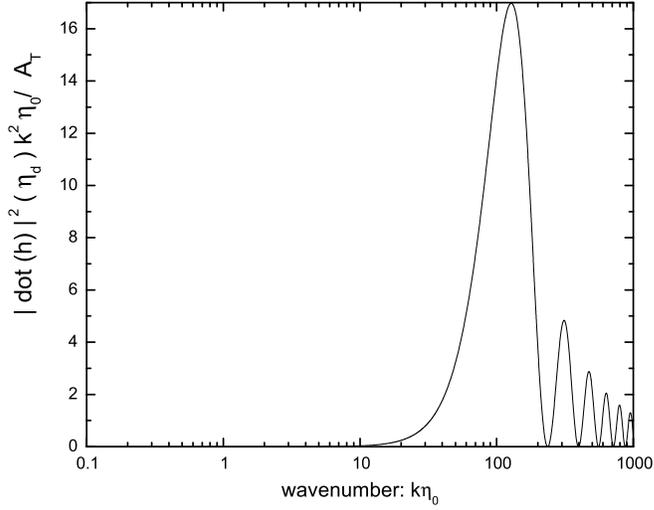}} \caption{\small
In the WKB approximation  the factor function
$|\dot{h}(\eta_d)|^2k^2\eta_0/A_{T}$ is peaked around $\sim 127$,
validating the relation (\ref{l}) as a fairly good estimate. Here
the factor $|\dot{h}(\eta_d)|^2k^2$ is multiplied  by a factor
$\eta_0/A_{T}$  only for a clear graphical demonstration. }
\end{figure}

%\begin{figure}
%\centerline{\includegraphics[width=12cm]{l_h.eps}}
%\caption{\small The influence of the dark energy $\Omega_{\Lambda}$ on
%the GW $h(\eta_d)$ at the decoupling.
%A larger value of $\Omega_{\Lambda}$ shifts the peaks of $h(\eta_d)$ to
%smaller scales. }
% \end{figure}

\begin{figure}
\centerline{\includegraphics[width=12cm]{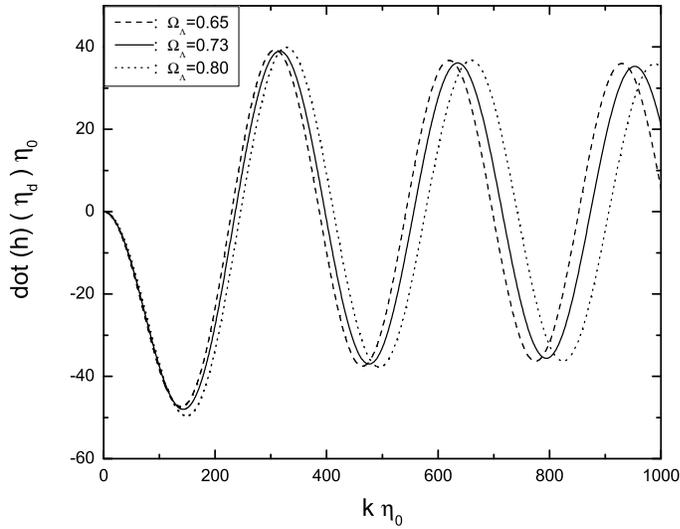}} \caption{\small
The influence of the dark energy $\Omega_{\Lambda}$ on the time
derivative of GW,  $\dot h(\eta_d)$,  at the decoupling. A larger
value of $\Omega_{\Lambda}$ shifts the peaks of $\dot h(\eta_d)$
to smaller scales. }
\end{figure}

\begin{figure}
\centerline{\includegraphics[width=12cm]{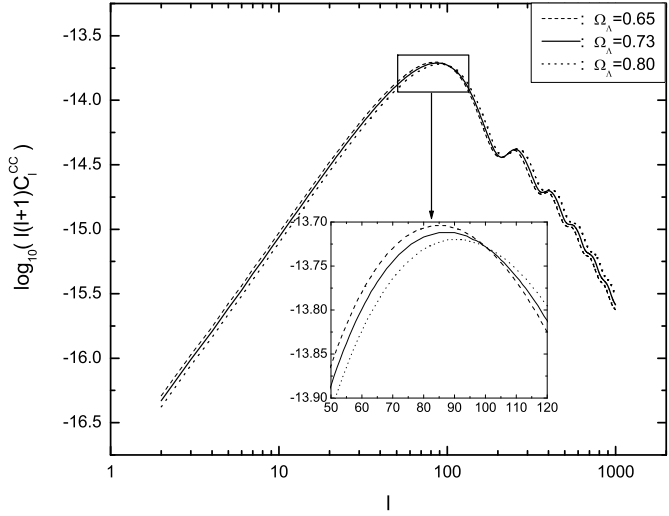}} \caption{\small
The magnetic polarization spectrum $C_l^{CC}$ depends weakly on
$\Omega_{\Lambda}$, and a larger value of $\Omega_{\Lambda}$
shifts the peaks of slightly to  smaller scales.}
\end{figure}

\begin{figure} \centerline{\includegraphics[width=12cm]{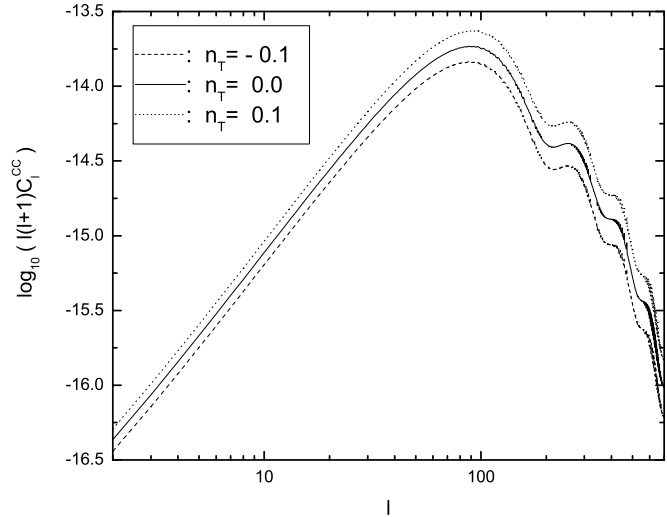}}
\caption{\small  The magnetic polarization spectrum $C_l^{CC}$
depends on the spectrum index $n_T$ of the relic GW.
$C_l^{CC}$ are plotted for three values of $n_T=-0.1$, $0.0$,
and $0.1$, respectively.
The following parameters are taken:
$r=1$, $\alpha =2$ in the half-gaussian model.
A larger value of $n_T$ produces
a higher spectrum $C_l^{CC}$.
}
\end{figure}

%\begin{figure} \centerline{\includegraphics[width=12cm]{re.eps}}
%\caption{\small We plot the optical depth function $\kappa(\eta)$
%and visibility function $V(\eta)\eta_0$ of the reionization
%process, where we have assumed instantaneous reionization with the
%optical depth $\kappa_r=0.09$. }
%\end{figure}

\end{document}